\documentclass{IEEEtran}
\usepackage{spconf,amsmath,graphicx,hyperref}
\usepackage[nolist,nohyperlinks]{acronym} 
\usepackage{algorithm,algpseudocode}
\usepackage{amsmath,amssymb,amsfonts}
\usepackage{graphicx,caption,subcaption,xspace,xcolor,cite}

\usepackage[labelfont=bf,font=small]{caption}
\def\BibTeX{{\rm B\kern-.05em{\sc i\kern-.025em b}\kern-.08em
    T\kern-.1667em\lower.7ex\hbox{E}\kern-.125emX}}

\usepackage{tikz,tikz-3dplot,tkz-tab}
\usetikzlibrary{angles, arrows.meta, calc, positioning, quotes, intersections,fit,decorations.pathreplacing,decorations.markings}
\tdplotsetmaincoords{70}{120}

\tikzset{
    cross/.style={fill=white,path picture={\draw[black]
        (path picture bounding box.south east) -- (path picture bounding box.north west)
        (path picture bounding box.south west) -- (path picture bounding box.north east);}},
    dressed/.style={fill=white,postaction={pattern=north east lines}},
    momentum/.style={->,semithick,yshift=5pt,shorten >=5pt,shorten <=5pt},
    loop/.style 2 args={thick,decoration={markings,mark=at position {#1} with {\arrow{<},\node[anchor=\pgfdecoratedangle-90,font=\footnotesize] {};}},postaction={decorate}},
    label/.style={thin,gray,shorten <=-1.5ex}
}

\tikzset{
    cross/.pic = {
    \draw[rotate = 45] (-#1,0) -- (#1,0);
    \draw[rotate = 45] (0,-#1) -- (0, #1);
    }
}

\begin{document}
    \begin{acronym}
        \acro{A-ALS}{adaptive alternating least squares}
        \acro{ALS}{alternating least squares}
        \acro{AoA}{azimuth angle of arrival}
        \acro{AoD}{azimuth angle of departure}
        \acro{ATTRACT}{adaptive tensor tracking for angular trajectories}
        \acro{AWGN}{additive white Gaussian noise}
        \acro{BALS}{bilinear alternating least squares}
        \acro{BD-RIS}{beyond-diagonal RIS}
        \acro{BTD}{block term decomposition}
        \acro{CRLB}{Cramér-Rao lower bound}
        \acro{CSI}{channel state information}
        \acro{DFRC}{dual-function radar-communication}
        \acro{DI}{Doppler ignorant}
        \acro{EoA}{elevation angle of arrival}
        \acro{EoD}{elevation angle of departure}
        \acro{ESPRIT}{estimation of signal parameters via rotational invariance techniques}
        \acro{EVD}{eigenvalue decomposition}
        \acro{FIM}{Fisher information matrix}
        \acro{HOSVD}{higher-order singular value decomposition}
        \acro{ISAC}{integrated sensing and communications}
        \acro{KF}{Kronecker factorization}
        \acro{KRSA}{Khatri-Rao sum approximation}
        \acro{KSA}{Kronecker sum approximation}
        \acro{LOS}{line-of-sight}
        \acro{LS}{least squares}
        \acro{MIMO}{multiple-input multiple-output}
        \acro{ML}{maximum likelihood}
        \acro{NLOS}{non-line-of-sight}
        \acro{NMSE}{normalized mean squared error}
        \acro{NTFE}{nested Tucker factorization estimation}
        \acro{OFDM}{orthogonal frequency division multiplexing}
        \acro{OMP}{orthogonal matching pursuit}
        \acro{PARAFAC}{parallel factor}
        \acro{PDF}{probability density function}
        \acro{RCS}{radar cross section}
        \acro{RIS}{reconfigurable intelligent surface}
        \acro{RMSE}{root mean squared error}
        \acro{SNR}{signal-to-noise ratio}
        \acro{SR}{sensing receiver}
        \acro{ST}{sensing transmitter}
        \acro{SVD}{singular value decomposition}
        \acro{TenDAE}{tensor Doppler-delay and angle estimation}
    \end{acronym}

    \title{PARAFAC-Based Low-Latency Angular Tracking for RIS-Assisted Sensing}
    
    \makeatletter
    \gdef\@name{\textit{Kenneth Ben\'{\i}cio, Andr\'{e} L. F. de Almeida, Bruno Sokal, Fazal-E Asim,} \\ \textit{Gabor Fodor, and A.~Lee Swindlehurst}}
    \makeatother
    
    \address{}
    
    \maketitle
    
    \begin{abstract}
        This paper proposes adaptive tensor tracking for angular trajectories (ATTRACT), a structure-aware alternating least squares algorithm for low-latency target tracking in reconfigurable intelligent surface (RIS)-assisted sensing. The received echo is represented as a dynamic third-order PARAFAC tensor that separates angular, delay, and Doppler information. Unlike batch tensor estimators, ATTRACT processes one data slice at a time and carries forward the preceding factor estimates, avoiding the need to collect an entire sensing window before updating the target state. Its distinguishing feature is the integration of delay and Doppler structure into each iterative update through the known transmitter--RIS channel, pilot signals, and RIS configurations. Numerical results show that, for a large enough RIS, ATTRACT achieves accuracy comparable to the offline tensor baseline at high signal-to-noise ratios while reducing computational overhead and providing slot-wise estimates.
    \end{abstract}
    

    \section{Introduction}
        \Ac{ISAC} enables base stations to serve users and sense their surroundings simultaneously \cite{liu2022integrated,zhang2025intelligent}. \Acp{RIS} support \ac{ISAC} by mitigating blockages and providing additional spatial degrees of freedom for high-resolution and extended sensing \cite{benicio2026multi, chepuri2023integrated, de2021channel,trice2021}. Reliable sensing nevertheless depends on accurate channel parameter estimation and suitable training protocols \cite{gong2020toward, Swindle2022, benicio2023tensor}. Tensor methods exploit the multidimensional structure of the received signal to formulate these tasks as structured tensor fitting problems.

        Our previous work in \cite{benicio2024low,benicio2024ris} studied estimation of the delay, Doppler, and angular parameters of a target using the nested \ac{PARAFAC} model in \cite{comon2009tensor}, solved using \ac{ALS} and \ac{ESPRIT}. Despite their accuracy, these batch frameworks assemble pilot blocks over a sensing window before estimation begins. For moving targets, this data-collection requirement delays parameter updates, while processing the accumulated observations increases the computational burden. Streaming tensor algorithms address this limitation through sequential updates \cite{abed2022contemporary,nion2009adaptive}. The contribution considered here is to combine this sequential processing feature with the physical factor structure of \ac{RIS}-assisted sensing, instead of relying solely on tensor updates that disregard this structure.

        To this end, we propose \ac{ATTRACT}, a structure-aware, slice-wise \ac{ALS} algorithm. Its main contributions are threefold. First, a dynamic third-order \ac{PARAFAC} representation separates the time-varying angular response from the delay- and Doppler-dependent factors, enabling updates from each newly available data slice. Second, the algorithm uses the known \ac{ST}--\ac{RIS} channel, pilot matrix, and \ac{RIS} configurations to refine the delay and Doppler factors within each iteration. This integration of recursive updates and sensing-specific structure is the central feature of \ac{ATTRACT}. Third, numerical comparisons with an offline \ac{ALS} baseline assess estimation accuracy, angular tracking, and execution time. The results show that \ac{ATTRACT} approaches the baseline accuracy at high \ac{SNR} while providing low-cost, slot-wise updates without waiting for the full observation window. \textit{Notations}: Scalars, vectors, matrices, and tensors are denoted by $a$, $\boldsymbol{a}$, $\boldsymbol{A}$, and $\boldsymbol{\mathcal{A}}$, respectively. The operators $(\cdot)^{\text{T}}$ and $(\cdot)^{\dagger}$ denote the transpose and pseudoinverse. The operator $\text{D}(\cdot)$ maps a vector to a diagonal matrix, and $\boldsymbol{I}_{N}$ is the $N \times N$ identity matrix. The $n$-mode product is $\boldsymbol{\mathcal{A}} \times_{n} \boldsymbol{B} = \boldsymbol{B} [\boldsymbol{\mathcal{A}}]_{(n)}$, while $\otimes$ and $\circ$ denote the Kronecker and Khatri-Rao products. 
    \section{System Model}
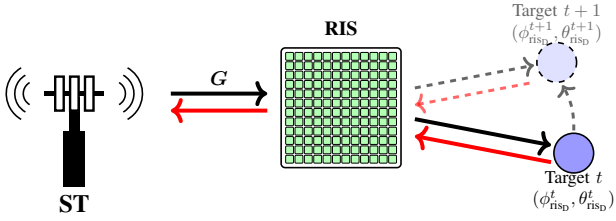
\begin{figure}[!t]
            \centering
            \resizebox{0.975\columnwidth}{!}{
            \begin{tikzpicture}
                \colorlet{risboard}{white}
                \colorlet{riselement}{green!30}
                \colorlet{txcolor}{black}
                \colorlet{rxcolor}{red}
                \begin{scope}[shift={(-7.5,-1.0)}, scale=0.75]
                    \fill (-0.25, 0) rectangle (0.25, 1.3);
                    \fill (-0.12, 1.3) rectangle (0.12, 2.1);
                    \fill (-0.7, 2.15) rectangle (0.7, 2.25);
                    \draw[very thick, fill=white] (-0.45, 1.8) rectangle (-0.25, 2.6);
                    \draw[very thick, fill=white] (-0.10, 1.8) rectangle (0.10, 2.6);
                    \draw[very thick, fill=white] (0.25, 1.8) rectangle (0.45, 2.6);
                    \foreach \r in {0.3, 0.5, 0.7} {
                        \draw[thick] (-0.9, 2.2) ++(120:\r) arc (120:240:\r);
                    }
                    \foreach \r in {0.3, 0.5, 0.7} {
                        \draw[thick] (0.9, 2.2) ++(60:\r) arc (60:-60:\r);
                    }
                    \node[font=\bfseries, scale=1.25] at (0, -0.4) {ST};
                \end{scope}
                \begin{scope}[shift={(-2.8, 0.4)}, scale=0.66]
                    \draw[thick, fill=risboard, rounded corners=2pt] (-1.6, -1.6) rectangle (1.6, 1.6);
                    \foreach \x in {-1.375, -1.125, ..., 1.375} {
                        \foreach \y in {-1.375, -1.125, ..., 1.375} {
                            \draw[line width=0.3pt, fill=riselement, rounded corners=0.5pt] (\x-0.105, \y-0.105) rectangle (\x+0.105, \y+0.105);
                        }
                    }
                    \node[above=0.1cm of {(0, 1.6)}, font=\bfseries] {RIS};
                \end{scope}
                \coordinate (T1) at (1.3, -0.4); 
                \coordinate (T2) at (1.0, 1.2);
                \fill[blue!40, draw=black, thick] (T1) circle (0.35);
                \node[below=0.2cm of T1, align=center, scale=0.9] {Target $t$ \\ $(\phi^{t}_{\text{ris}_{\text{D}}}, \theta^{t}_{\text{ris}_{\text{D}}})$};
                \fill[blue!40, draw=black, thick, dashed, fill opacity=0.3] (T2) circle (0.35);
                \node[above=0.2cm of T2, align=center, text=gray!80!black, scale=0.9] {Target $t+1$ \\ $(\phi^{t+1}_{\text{ris}_{\text{D}}}, \theta^{t+1}_{\text{ris}_{\text{D}}})$};
                \draw[->, line width=1.5pt, dashed, gray!80!black] (1.3, 0.0) to[bend right=15] (1.1, 0.8);
                \draw[->, line width=2pt, txcolor] (-5.8, 0.65) -- node[above] {$\boldsymbol{G}$} (-4.1, 0.65);
                \draw[<-, line width=2pt, rxcolor] (-5.8, 0.35) -- (-4.1, 0.35);  
                \draw[->, line width=2pt, txcolor] (-1.5, 0.2) -- (0.9, -0.26); 
                \draw[<-, line width=2pt, rxcolor] (-1.5, -0.1) -- (0.9, -0.56);
                \draw[->, line width=1.5pt, txcolor, dashed, opacity=0.6] (-1.5, 0.75) -- (0.6, 1.15);
                \draw[<-, line width=1.5pt, rxcolor, dashed, opacity=0.6] (-1.5, 0.45) -- (0.6, 0.85);
         
            \end{tikzpicture}
            }
            \caption{RIS-assisted sensing of a moving target with time-varying departure angles.}
            \label{fig:system_model_localization}
        \end{figure}

        Consider the scenario in Fig.~\ref{fig:system_model_localization}, where an \ac{ST} with a uniform planar array of $L = L_{z} L_{y}$ elements is assisted by a diagonal \ac{RIS} composed of $N = N_{z} N_{y}$ elements with unit-modulus response. The direct \ac{LOS} path to the moving target is assumed to be blocked, so sensing relies exclusively on the \ac{RIS}-assisted link. The \ac{ST} transmits pilot vectors $\boldsymbol{x}_q \in \mathbb{C}^{L\times 1}$ toward the \ac{RIS} using an \ac{OFDM} waveform comprising $Q$ subcarriers and $M$ symbols, and receives the target echo via the same RIS. The received echo at the \ac{ST} is given by
        \begin{align}
            \nonumber \boldsymbol{y}_{q,m,t} &= \underbrace{\boldsymbol{G} \text{D}(\boldsymbol{w}_{m}) \boldsymbol{p}(\phi^{t}_{\text{ris}_{\text{D}}}, \theta^{t}_{\text{ris}_{\text{D}}})}_{\text{Target-RIS-ST path}} \underbrace{\boldsymbol{p}^{\text{T}}(\phi^{t}_{\text{ris}_{\text{D}}}, \theta^{t}_{\text{ris}_{\text{D}}}) \text{D}(\boldsymbol{w}_{m}) \boldsymbol{G}^{\text{T}}}_{\text{ST-RIS-Target path}} \\
            &\times \boldsymbol{x}_{q} [\boldsymbol{c}(\tau)]_{q} [\boldsymbol{d}(\nu)]_{m} \alpha_{t} + \boldsymbol{z}_{q,m,t},
        \end{align}
        \textcolor{black}{where $\boldsymbol{G} \in \mathbb{C}^{L \times N}$ models the geometric \ac{ST}-\ac{RIS} channel in a rich scattering environment, i.e., $\boldsymbol{G}$ is assumed to be a full-rank channel matrix.} The steering vector from the \ac{RIS} toward the target at time slot $t$ is $\boldsymbol{p}(\phi^{t}_{\text{ris}_{\text{D}}}, \theta^{t}_{\text{ris}_{\text{D}}}) \in \mathbb{C}^{N \times 1}$, where $\phi^{t}_{\text{ris}_{\text{D}}}$ and $\theta^{t}_{\text{ris}_{\text{D}}}$ are the time-varying \ac{AoD} and \ac{EoD}, respectively. \textcolor{black}{The \ac{RIS} configuration vector is $\boldsymbol{w}_{m} \in \mathbb{C}^{N \times 1}$}. With half-wavelength element spacing and spatial frequencies $\mu_{\text{st}} = \pi \text{sin}(\phi_{\text{st}}) \text{sin}(\theta_{\text{st}})$ and $\psi_{\text{st}} = \pi \text{cos}(\phi_{\text{st}})$, the \ac{ST} array response in the $y$--$z$ plane is \cite{Asim_2025}
        \begin{align}
            \boldsymbol{a}(\mu_{\text{st}},\psi_{\text{st}}) = \boldsymbol{a}_{y}(\mu_{\text{st}}) \otimes \boldsymbol{a}_{z}(\psi_{\text{st}}) \in \mathbb{C}^{L \times 1},
        \end{align}
        where $\boldsymbol{a}_{y}(\mu_{\text{st}}) = [1,  e^{-j\mu_{\text{st}}}, \cdots, e^{-j  (L_{y} - 1) \mu_{\text{st}}} ]^{\text{T}} \in \mathbb{C}^{L_{y} \times 1}$, and $\boldsymbol{a}_{z}(\psi_{\text{s}t}) = [1,  e^{-j\psi_{\text{st}}}, \cdots, e^{-j  (L_{z} - 1) \psi_{\text{st}}} ]^{\text{T}} \in \mathbb{C}^{L_{z} \times 1}$.
        The \ac{RIS} array responses $\boldsymbol{b}(\phi_{\text{ris}_{A}},\theta_{\text{ris}_{A}})$ and $\boldsymbol{p}(\phi^{t}_{\text{ris}_{\text{D}}}, \theta^{t}_{\text{ris}_{\text{D}}})$ are defined analogously. 
        The target delay $\tau$ gives rise to the frequency-domain steering vector $\boldsymbol{c}(\tau) = [1, \cdots, e^{-j 2 \pi (Q - 1) \Delta f \tau}]^{\text{T}} \in \mathbb{C}^{Q \times 1}$, whose $q$th entry weights the $q$th \ac{OFDM} subcarrier. The Doppler shift $\nu$ induces a phase progression across symbols: $\boldsymbol{d}(\nu) = [1, \cdots, e^{j 2 \pi (M - 1) T_{s} \nu}]^{\text{T}} \in \mathbb{C}^{M \times 1}$. The noise vector at subcarrier $q$, symbol $m$, and time slot $t$ is $\boldsymbol{z}_{q,m,t} \in \mathbb{C}^{L \times 1}$. The channel coefficient $\alpha_t \in \mathbb{C}$ accounts for path loss and \ac{RCS} fluctuations, following \cite{benicio2026multi,ercan2025ris}. 
        
        Stacking the samples across the $Q$ subcarriers for a fixed symbol and time slot gives the following noiseless echo matrix:
        \vspace{-1ex}
        \begin{align}
            \notag \boldsymbol{Y}_{m,t} &= [\boldsymbol{y}_{1,m}, \cdots, \boldsymbol{y}_{Q,m}] \in \mathbb{C}^{L \times Q}, \\
            \notag &= \boldsymbol{G} \text{D}(\boldsymbol{w}_{m}) \boldsymbol{P}_{t} \text{D}(\boldsymbol{w}_{m}) \boldsymbol{G}^{\text{T}} \boldsymbol{X} \text{D}(\boldsymbol{c}(\tau)) [\boldsymbol{d}(\nu)]_{m}.
        \end{align}
        \textcolor{black}{where $\boldsymbol{X} \in \mathbb{C}^{L \times Q}$ is the transmitted pilot matrix}, and $\boldsymbol{P}_{t} = \alpha_{t} \boldsymbol{p}(\boldsymbol{\phi^{t}_{\text{ris}_{\text{D}}}, \theta^{t}_{\text{ris}_{\text{D}}}}) \boldsymbol{p}^{\text{T}}(\boldsymbol{\phi^{t}_{\text{ris}_{\text{D}}}, \theta^{t}_{\text{ris}_{\text{D}}}}) \in \mathbb{C}^{N \times N}$ combines the target's time-varying angular response and complex channel coefficient. We omit the additive noise term in the following derivation for clarity, but retain it in the statistical model and numerical evaluations. Using the identities $\text{vec}(\boldsymbol{A} \boldsymbol{B} \boldsymbol{C}) = (\boldsymbol{C}^{\text{T}} \otimes \boldsymbol{A}) \text{vec}(\boldsymbol{B})$ and $\text{vec}(\boldsymbol{A} \text{D}(\boldsymbol{b}) \boldsymbol{C}) = (\boldsymbol{C}^{\text{T}} \diamond \boldsymbol{A}) \boldsymbol{b}$, together with $\boldsymbol{a}^{\text{T}} \diamond \boldsymbol{B} = \boldsymbol{B} \text{D}(\boldsymbol{a})$, the echo signal associated with the $m$th symbol can be written as
        \vspace{-1ex}
        \begin{align}
            \notag \boldsymbol{y}_{m,t} &\hspace{-0.1cm}=\hspace{-0.1cm} [(\text{D}(\boldsymbol{c}(\tau)) \boldsymbol{X}^{\text{T}} \boldsymbol{G}) \otimes \boldsymbol{G}] \text{D}(\text{vec}(\boldsymbol{P}_{t})) (\boldsymbol{w}_{m} \hspace{-0.05cm}\otimes\hspace{-0.05cm} \boldsymbol{w}_{m}) [\boldsymbol{d}(\nu)]_{m}.
        \end{align}
        Stacking the vectors for all $M$ \ac{OFDM} symbols as columns of $\boldsymbol{\mathcal{Y}}_{..t} \in \mathbb{C}^{L Q \times M}$ yields
         \vspace{-1ex}
        \begin{align}
            \boldsymbol{\mathcal{Y}}_{..t} &= \boldsymbol{X}_{\tau} \text{D}(\boldsymbol{p}_{t}) \boldsymbol{W}_{\nu} \in \mathbb{C}^{L Q \times M}, \label{eq:frontal_slices}
        \end{align}
        where $\boldsymbol{W} \in \mathbb{C}^{N \times M}$ collects the \ac{RIS} configurations over the $M$ symbols, $\boldsymbol{X}_{\tau} = [\boldsymbol{A}_{\tau} \otimes \boldsymbol{G}] \in \mathbb{C}^{L Q \times N^{2}}$ is the delay-dependent factor, $\boldsymbol{A}_{\tau} = \text{D}(\boldsymbol{c}(\tau)) \boldsymbol{X}^{\text{T}} \boldsymbol{G} \in \mathbb{C}^{Q \times N}$, $\boldsymbol{W}_{\nu} = (\boldsymbol{W} \diamond \boldsymbol{W}) \text{D}(\boldsymbol{d}(\nu)) \in \mathbb{C}^{N^{2} \times M}$ is the Doppler-dependent factor, and $\boldsymbol{p}_{t} = \text{vec}(\boldsymbol{P}_{t}) \in \mathbb{C}^{N^2 \times 1}$ contains the time-varying angular response and complex channel coefficient. 
        
        Stacking the matrices in (\ref{eq:frontal_slices}) along the temporal dimension yields a third-order \ac{PARAFAC} tensor $\boldsymbol{\mathcal{Y}} \in \mathbb{C}^{L Q \times M \times T(t)}$, expressed as \cite{nion2009adaptive,abed2022contemporary}
         \vspace{-0ex}
        \begin{align}
            \boldsymbol{\mathcal{Y}} = \boldsymbol{\mathcal{I}}_{3,N^{2}} \times_{1} \boldsymbol{X}_{\tau} \times_{2} \boldsymbol{W}^{\text{T}}_{\nu} \times_{3} \boldsymbol{P}', \label{eq:PARAFAC}
        \end{align}
        whose global unfoldings are given by
         \vspace{-.1ex}
        \begin{align}
            [\boldsymbol{\mathcal{Y}}]_{(1)} &= \boldsymbol{X}_{\tau} (\boldsymbol{P}' \diamond \boldsymbol{W}^{\text{T}}_{\nu})^{\text{T}} \in \mathbb{C}^{L Q \times M T(t)}, \label{eq:tensor_unfolding_1} \\
            [\boldsymbol{\mathcal{Y}}]_{(2)} &= \boldsymbol{W}^{\text{T}}_{\nu} (\boldsymbol{P}' \diamond \boldsymbol{X}_{\tau})^{\text{T}} \in \mathbb{C}^{M \times L Q T(t)}, \label{eq:tensor_unfolding_2} \\
            [\boldsymbol{\mathcal{Y}}]_{(3)} &= \boldsymbol{P}' (\boldsymbol{W}^{\text{T}}_{\nu} \diamond \boldsymbol{X}_{\tau})^{\text{T}} \in \mathbb{C}^{T(t) \times L Q M}, \label{eq:tensor_unfolding_3}
        \end{align}
        where $\boldsymbol{P}' = [\boldsymbol{p}_{1}, \cdots, \boldsymbol{p}_{T}]^{\text{T}} \in \mathbb{C}^{T(t) \times N^{2}}$ collects the angular factors available at time $t$. This representation separates the evolving target response from the delay and Doppler factors, which are treated as static over the observation window. Rather than repeatedly fitting the accumulated tensor, the proposed algorithm updates these factors recursively and estimates the angular state from the current slice.

    \section{PARAFAC-Based Adaptive Tensor Tracking for Angular Trajectories}
        We fit the \ac{PARAFAC} model in (\ref{eq:frontal_slices}) using adaptive \ac{ALS} \cite{nion2009adaptive} to estimate $\boldsymbol{p}_t$, $\boldsymbol{c}(\tau)$, and $\boldsymbol{d}(\nu)$. Each newly available data slice is processed using the preceding factor estimates, and the known sensing structure is incorporated within the iterative updates. After convergence at time $t$, \ac{ESPRIT} extracts the physical parameters, and the complex channel coefficient is estimated from the reconstructed echo. 
        
        Using (\ref{eq:frontal_slices}), we formulate the dynamic fitting problem as
        \vspace{-1.2ex}
        \begin{align}
            \notag \{\hat{\boldsymbol{X}}_{\tau, t}, \hat{\boldsymbol{W}}_{\nu, t}, \hat{\boldsymbol{p}}_{t}\} = \underset{\boldsymbol{X}_{\tau, t-1}, \boldsymbol{W}_{\nu, t-1}, \boldsymbol{p}_{t-1}}{\text{arg min}} \left\| \begin{aligned} \boldsymbol{\mathcal{Y}}_{..t} - \hat{\boldsymbol{X}}_{\tau,t-1} \\ \text{D}(\hat{\boldsymbol{p}}_{t-1}) \hat{\boldsymbol{W}}_{\nu,t-1}\end{aligned} \right\|_{\text{F}}^{2}. \label{eq:tensor_fit_problem}
        \end{align}
        Adaptive \ac{ALS} solves this problem through sequential \ac{LS} updates of the factors associated with the current slice. To obtain the angular update, we first vectorize the $t$th data slice $\boldsymbol{\mathcal{Y}}_{..t}$ as
        \begin{align}
            \boldsymbol{y}_{t} = \text{vec}(\boldsymbol{\mathcal{Y}}_{..t}) = (\boldsymbol{W}_{\nu}^{\text{T}} \diamond \boldsymbol{X}_{\tau}) \boldsymbol{p}_{t}.
        \end{align}
        Although (\ref{eq:tensor_unfolding_1})--(\ref{eq:tensor_unfolding_3}) describe the global tensor unfoldings, \ac{ATTRACT} operates only on the two-dimensional data slice at time $t$. The angular update uses the vectorized observation $\boldsymbol{y}_t$, whereas the delay and Doppler updates use the matrix $\boldsymbol{\mathcal{Y}}_{..t}$. \textcolor{black}{Let $i$ denote the inner \ac{ALS} iteration index. At time $t$, the variables are initialized using the final estimates from $t-1$: $\hat{X}_{\tau,t}^{i} = \hat{X}_{\tau,t-1}$, $\hat{W}_{\nu,t}^{i} = \hat{W}_{\nu,t-1}$, and $\hat{p}_{t}^{i} = \hat{p}_{t-1}$, for $i = 0$.} Starting from the preceding estimates $\hat{\boldsymbol{X}}_{\tau, t-1}$ and $\hat{\boldsymbol{W}}_{\nu, t-1}$, the factors are updated through the following \ac{LS} subproblems until a stopping criterion is met\footnote{As specified in Alg.~\ref{alg:proposed_tracking}, the iterations stop when the change in the objective function falls below the prescribed tolerance or the maximum number of iterations is reached.}:
        \textcolor{black}{\begin{align}
            \hat{\boldsymbol{p}}^{i}_{t} &= \underset{\boldsymbol{p}_{t}}{\text{arg min}} \left\| \boldsymbol{y}_{t} - ((\boldsymbol{W}^{i-1}_{\nu, t})^{\text{T}} \diamond \boldsymbol{X}^{i-1}_{\tau, t}) \boldsymbol{p}^{i}_{t} \right\|_{2}^{2}, \\
            \hat{\boldsymbol{X}}^{i}_{\tau,t} &= \underset{\boldsymbol{X}_{\tau,t}}{\text{arg min}}  \left\| \boldsymbol{\mathcal{Y}}_{..t} - \boldsymbol{X}^{i}_{\tau,t} \text{D}(\hat{\boldsymbol{p}}^{i}_{t}) \boldsymbol{W}^{i-1}_{\nu,t} \right\|_{\text{F}}^{2}, \\
            \hat{\boldsymbol{W}}^{i}_{\nu,t} &= \underset{\hat{\boldsymbol{W}}_{\nu,t}}{\text{arg min}}  \left\| \boldsymbol{\mathcal{Y}}_{..t} - \hat{\boldsymbol{X}}^{i}_{\tau,t}  \text{D}(\hat{\boldsymbol{p}}^{i}_{t}) \boldsymbol{W}^{i}_{\nu,t} \right\|_{\text{F}}^{2}.
        \end{align}
        These subproblems yield the following slot-wise updates:
        \begin{align}
            \hat{\boldsymbol{p}}^{i}_{t} &= ((\hat{\boldsymbol{W}}^{i-1}_{\nu, t})^{\text{T}} \diamond \hat{\boldsymbol{X}}^{i-1}_{\tau, t})^{\dagger} \boldsymbol{y}_{t}, \label{eq:est_angle} \\
            \hat{\boldsymbol{X}}^{i}_{\tau,t} &= \boldsymbol{\mathcal{Y}}_{..t} (\text{D}(\hat{\boldsymbol{p}}^{i}_{t}) \hat{\boldsymbol{W}}^{i-1}_{\nu,t})^{\dagger}, \label{eq:est_X_delay}\\
            \hat{\boldsymbol{W}}^{i}_{\nu,t} &= (\hat{\boldsymbol{X}}^{i}_{\tau,t}  \text{D}(\hat{\boldsymbol{p}}^{i}_{t}))^{\dagger} \boldsymbol{\mathcal{Y}}_{..t}, \label{eq:est_W_doppler}
        \end{align}}
        \textcolor{black}{which provide initial factor estimates for the current slice, provided $LMQ \geq N^{2}$, $M \geq N$, and $LQ \geq N^{2}$, respectively.} The key refinement in \ac{ATTRACT} is to enforce the sensing-specific structure before the next \ac{ALS} iteration, rather than defer all physical interpretation until after convergence. Specifically, the internal structure of $\hat{\boldsymbol{X}}_{\tau,t}$ and $\hat{\boldsymbol{W}}_{\nu,t}$ is used to estimate the delay vector $\boldsymbol{c}(\tau)$ and Doppler vector $\boldsymbol{d}(\nu)$ and reconstruct the corresponding factors within the iterative estimation loop.
        
        For delay refinement, we assume that the \ac{ST}--\ac{RIS} channel matrix $\boldsymbol{G}$ is available from a prior estimation procedure \cite{benicio2023tensor}. \textcolor{black}{Note that the unstructured update in (\ref{eq:est_X_delay}) cannot uniquely recover the full Kronecker structure of $\boldsymbol{X}_{\tau}$, since the $N^{2} - N$ cross-column terms ($\boldsymbol{a}_{i} \otimes \boldsymbol{g}_{j}, i \neq j$) are unrecoverable\footnote{The commutativity in $\boldsymbol{W}_{\nu,t} \propto \boldsymbol{W} \diamond \boldsymbol{W}$ yields identical rows. This inherent rank deficiency prevents a unique estimation of all columns of $\hat{\boldsymbol{X}}_{\tau,t}$ in problem (\ref{eq:est_X_delay}).} However, the update in (\ref{eq:est_X_delay}) successfully resolves the $N$ matched columns ($\boldsymbol{a}_n \otimes \boldsymbol{g}_n$), whose indices are $i_{n} = (n - 1) N + n$, for $n \in \{1, \dots, N\}$, corresponding to the Khatri-Rao structure $\tilde{\boldsymbol{X}}_{\tau} = \boldsymbol{A}_{\tau} \diamond \boldsymbol{G} \in \mathbb{C}^{L Q \times N}$ whose $n$th column satisfies $\tilde{\boldsymbol{x}}_{\tau,n} = \boldsymbol{a}_{\tau,n} \otimes \boldsymbol{g}_{n} = \text{vec}(\boldsymbol{g}_{n} \boldsymbol{a}^{\text{T}}_{\tau,n})$, leading to the estimate}
        \begin{align*}
            \hat{\boldsymbol{a}}_{\tau,n} = (\boldsymbol{I}_{Q} \otimes \boldsymbol{g}_{n})^{\dagger} \hat{\boldsymbol{x}}_{\tau,n}, \forall n \in \{1, \cdots, N\}.
        \end{align*}
        After assembling $\hat{\boldsymbol{A}}_{\tau}=[\hat{\boldsymbol{a}}_{\tau,1}, \ldots, \hat{\boldsymbol{a}}_{\tau,N}]$, the delay vector is estimated as
         \vspace{-1ex}
        \begin{align}
            \hat{\boldsymbol{c}}(\tau) = \left( (\boldsymbol{G}^T\boldsymbol{X}) \diamond \boldsymbol{I}_Q \right)^\dagger \text{vec}(\hat{\boldsymbol{A}}_{\tau}). \label{eq:delay_subspace}
        \end{align}
        
        Similarly, the known \ac{RIS} configuration $\boldsymbol{W}$ is used to refine the Doppler factor within the \ac{ALS} loop. The structured relationship is $\text{vec}(\boldsymbol{W}_{\nu}) = [\boldsymbol{I}_{M} \diamond (\boldsymbol{W} \diamond \boldsymbol{W})] \boldsymbol{d}(\nu)$,which gives the following \ac{LS} estimate of the Doppler vector from $\hat{\boldsymbol{W}}_{\nu,t}$:
        \begin{align}
            \hat{\boldsymbol{d}}(\nu) = [\boldsymbol{I}_{M} \diamond (\boldsymbol{W} \diamond \boldsymbol{W})]^{\dagger} \text{vec}(\hat{\boldsymbol{W}}_{\nu,t}). \label{eq:doppler_subspace}
        \end{align} 
        At the end of each iteration, the Doppler factor is reconstructed as $\hat{\boldsymbol{W}}_{\nu,t} = (\boldsymbol{W} \diamond \boldsymbol{W}) \text{D}(\hat{\boldsymbol{d}}(\nu))$. Together with the delay-factor reconstruction in Alg.~\ref{alg:proposed_tracking}, this step restricts the updates to the known sensing structure instead of treating all factor entries as independent unknowns. After \ac{ALS} converges, \ac{ESPRIT} uses $\hat{\boldsymbol{c}}(\tau)$ and $\hat{\boldsymbol{d}}(\nu)$ to estimate $\hat{\tau}$ and $\hat{\nu}$, respectively. \textcolor{black}{Finally, to estimate the time-varying complex channel coefficient, we parametrically reconstruct the angular, delay, and Doppler factors using the parameters estimated by \ac{ESPRIT} to obtain scale-free estimates. Thus, the complex channel gain is extracted by computing the sample mean of the element-wise ratio between the received data slice and the reconstructed slice
        \begin{align}
            \hat{\alpha}_t = \frac{1}{LQM} \sum_{i=1}^{LQ} \sum_{j=1}^{M} \left[ \mathcal{Y}_{..t} \oslash \tilde{\mathcal{Y}}_{..t} \right]_{i,j} \label{eq:complex_channel_gain_estimation}
        \end{align}
        where $[\cdot]_{i,j}$ denotes the $(i,j)$-th matrix element. This ratio effectively compensates for the spatial and temporal responses to recover the complex amplitude. Algorithm~\ref{alg:proposed_tracking} summarizes the procedure. Importantly, \ac{ATTRACT} does not require an explicit target-motion model; temporal information is retained through the preceding factor estimates, while the current observation updates the angular state.}
        \begin{algorithm}[!t]
            \caption{\Ac{ATTRACT}}
            \label{alg:proposed_tracking}
            \footnotesize
            \begin{algorithmic}[1]
                \Require{$\boldsymbol{\mathcal{Y}}_{..t}$, $\forall t \in \{1, \dots, T\}$, $\boldsymbol{G}$, $\boldsymbol{W}$, $i_{\text{max}}$, $\delta$.}
                \State{Randomly initialize $\hat{\boldsymbol{X}}_{\tau, 0}$, $\hat{\boldsymbol{W}}_{\nu, 0}$, and $\hat{\boldsymbol{p}}_{0}$}
                \For{$t = 1$ \textbf{to} $T$}
                    \State{At $i = 0$: $\hat{\boldsymbol{X}}_{\tau,t}^{0} = \hat{\boldsymbol{X}}_{\tau,t-1}$, $\hat{\boldsymbol{W}}_{\nu,t}^{0} = \hat{\boldsymbol{W}}_{\nu,t-1}$, and $\hat{\boldsymbol{p}}_{t}^{0} = \hat{\boldsymbol{p}}_{t-1}$}
                    \While{$||e(i) - e(i-1)|| \geq \delta$ \textbf{and} $i \leq i_{\text{max}}$}
                        \State{Update $\hat{\boldsymbol{p}}_{t}^{i} = (\hat{\boldsymbol{W}}_{\nu, t}^{(i-1)\text{T}} \diamond \hat{\boldsymbol{X}}_{\tau, t}^{i-1})^{\dagger} \boldsymbol{y}_{t}$}
                        \State{Update $\hat{\boldsymbol{X}}_{\tau,t}^{i} = \boldsymbol{\mathcal{Y}}_{..t} (\text{D}(\hat{\boldsymbol{p}}_{t}^{i}) \hat{\boldsymbol{W}}_{\nu,t}^{i-1})^{\dagger}$}
                        \For{$n = 1$ \textbf{to} $N$}
                            \State{Define $\tilde{\boldsymbol{x}}_{\tau,n} = \hat{\boldsymbol{X}}_{\tau,t}^{i}(:, (n-1)N + n)$}
                            \State{Estimate delay vector $\hat{\boldsymbol{a}}_{\tau,n} = (\boldsymbol{g}_{n}^{\dagger} \text{unvec}_{L \times Q}(\tilde{\boldsymbol{x}}_{\tau, n}))^{\text{T}}$}
                        \EndFor
                        \State{Build $\hat{\boldsymbol{A}} = [\hat{\boldsymbol{a}}_{\tau,1}, \dots, \hat{\boldsymbol{a}}_{\tau,N}]$}
                        \State{Estimate $\hat{\boldsymbol{c}}(\tau) = ((\boldsymbol{X}^{\text{T}} \boldsymbol{G})^{\text{T}} \diamond \boldsymbol{I}_Q)^{\dagger} \text{vec}(\hat{\boldsymbol{A}})$}
                        \State{Refine $\hat{\boldsymbol{X}}_{\tau,t}^{i} = (\text{D}(\hat{\boldsymbol{c}}(\tau)) \boldsymbol{X}^{\text{T}} \boldsymbol{G}) \otimes \boldsymbol{G}$}
                        \State{Update $\hat{\boldsymbol{W}}_{\nu,t}^{i} = (\hat{\boldsymbol{X}}_{\tau,t}^{i}  \text{D}(\hat{\boldsymbol{p}}_{t}^{i}))^{\dagger} \boldsymbol{\mathcal{Y}}_{..t}$}
                        \State{Estimate $\hat{\boldsymbol{d}}(\nu) = [\boldsymbol{I}_{M} \diamond (\boldsymbol{W} \diamond \boldsymbol{W})]^{\dagger} \text{vec}(\hat{\boldsymbol{W}}_{\nu,t}^{i})$}
                        \State{Refine $\hat{\boldsymbol{W}}_{\nu,t}^{i} = (\boldsymbol{W} \diamond \boldsymbol{W}) \text{D}(\hat{\boldsymbol{d}}(\nu))$}
                        \State{Update $e(i) = \| \boldsymbol{\mathcal{Y}}_{..t} - \hat{\boldsymbol{X}}_{\tau,t}^{i} \text{D}(\hat{\boldsymbol{p}}_{t}^{i}) \hat{\boldsymbol{W}}_{\nu,t}^{i} \|_{\text{F}}^{2}$ and $i \gets i + 1$}
                    \EndWhile
                    \State{Assign $\hat{\boldsymbol{p}}_{t} = \hat{\boldsymbol{p}}_{t}^{i-1}$, $\hat{\boldsymbol{X}}_{\tau,t} = \hat{\boldsymbol{X}}_{\tau,t}^{i-1}$, $\hat{\boldsymbol{W}}_{\nu,t} = \hat{\boldsymbol{W}}_{\nu,t}^{i-1}$}
                    \State{Extract (\ac{ESPRIT}) $\hat{\phi}^{t}_{\text{ris}_{\text{D}}}$, $\hat{\theta}^{t}_{\text{ris}_{\text{D}}}$, $\hat{\tau}$,  $\hat{\nu}$ from $\hat{\boldsymbol{p}}_{t}$, $\hat{\boldsymbol{c}}(\tau)$, and $\hat{\boldsymbol{d}}(\nu)$}
                    \State{Reconstruct $\hat{\boldsymbol{\mathcal{Y}}}_{..t} = \hat{\boldsymbol{X}}_{\tau,t} \text{D}(\hat{\boldsymbol{p}}_{t}) \hat{\boldsymbol{W}}_{\nu,t}$}
                    \State{Estimate $\hat{\alpha}_{t}$ with (\ref{eq:complex_channel_gain_estimation})}
                    \State{\textbf{return} $\hat{\phi}^{t}_{\text{ris}_{\text{D}}}$, $\hat{\theta}^{t}_{\text{ris}_{\text{D}}}$, $\hat{\nu}$, $\hat{\tau}$, and $\hat{\alpha}_{t}$}
                \EndFor
            \end{algorithmic}
        \end{algorithm}
    \vspace{-.2cm}  
    \section{Numerical Results}
        We evaluate \ac{ATTRACT} over $1000$ Monte Carlo runs in a far-field monostatic \ac{OFDM} scenario with a diagonal \ac{RIS} comprising $N = N_{y} \times N_{z}$ elements, \textcolor{black}{with $N_{y} = N_{z}$}, and random pilot signals. The performance metrics are $\text{RMSE}(\boldsymbol{x}) = \sqrt{\mathbb{E}\{||\boldsymbol{x} - \hat{\boldsymbol{x}}||^{2}_{2}\}}$ and \textcolor{black}{$\text{SNR} = ||\boldsymbol{\mathcal{Y}}||^{2}_{\text{F}}/||\boldsymbol{\mathcal{Z}}||^{2}_{\text{F}}$, where $\boldsymbol{\mathcal{Z}}$ is the \ac{AWGN} tensor component}, with delay and Doppler estimates normalized as $\tau/T_s$ and $\nu T_s$. The system operates at $28$~GHz ($\lambda = 1.07$~cm, $\Delta f = 120$~kHz), and the \ac{ST} uses a planar array with $L = 2 \times 2 = 4$ elements and half-wavelength spacing. The \ac{OFDM} configuration comprises $M = 256$ symbols, $Q = 16$ subcarriers, and $T = 32$ time slots. The \acs{ST}--\acs{RIS} and \acs{RIS}--target distances are drawn from $\mathcal{U}(10, 250)$~m. Initial target angles and relative velocities are drawn from $\mathcal{U}(10^{\circ}, 80^{\circ})$ and $\mathcal{U}(-10, 10)$~m/s, respectively, and the target \acs{RCS} is $2~\text{m}^{2}$. \textcolor{black}{The target angles nominally evolve according to a discrete random walk: $\phi_{t} = \phi_{t-1} + \delta_{1,t}$ and $\theta_{t} = \theta_{t-1} + \delta_{2,t}$, where $\delta_{1,t}$ and $\delta_{2,t}$ take the values $-0.1^{\circ}$ and $+0.1^{\circ}$ with equal probability. This model represents small slot-to-slot angular variations, as considered in channel-tracking studies \cite{zhang2016tracking}.} The \ac{ALS} iterations stop at $i_{\text{max}} = 100$ or when the change in the objective falls below $\delta = 10^{-3}$. We run simulations on a computer with an Intel i7-10700K processor and 32~GB of RAM.

\begin{figure}[!t]
            \centering
            \setlength{\abovecaptionskip}{3pt}
            \setlength{\belowcaptionskip}{-1pt}
            \begin{minipage}{0.475\columnwidth}
                \centering
                \includegraphics[width=\textwidth]{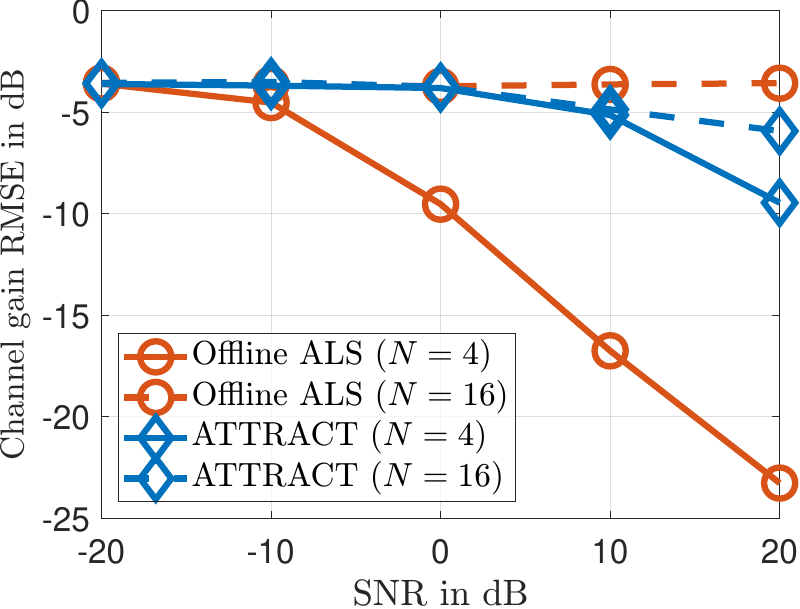}
                \caption*{(a) Channel gain.}  
            \end{minipage}
            \hfill
            \begin{minipage}{0.475\columnwidth}  
                \centering
                \includegraphics[width=\textwidth]{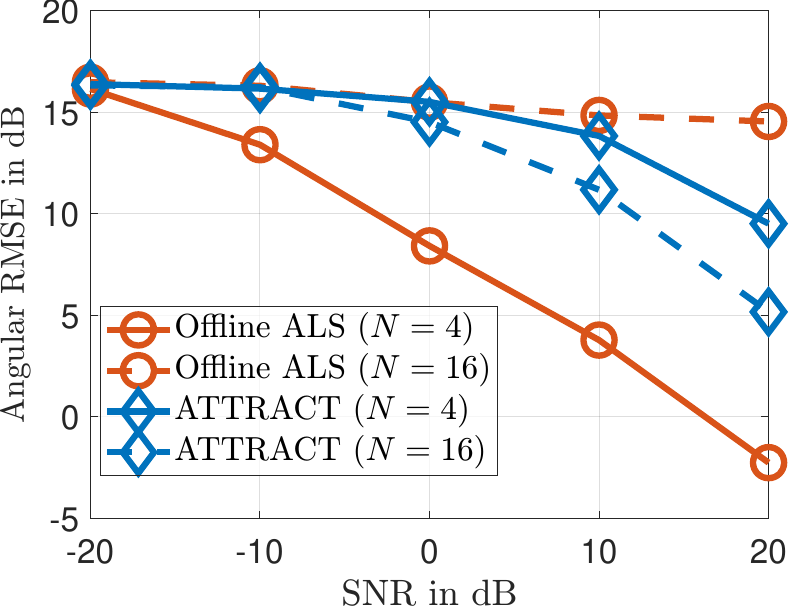}
                \caption*{(b) Angle.}   
            \end{minipage} 
            \begin{minipage}{0.475\columnwidth}   
                \centering
                \includegraphics[width=\textwidth]{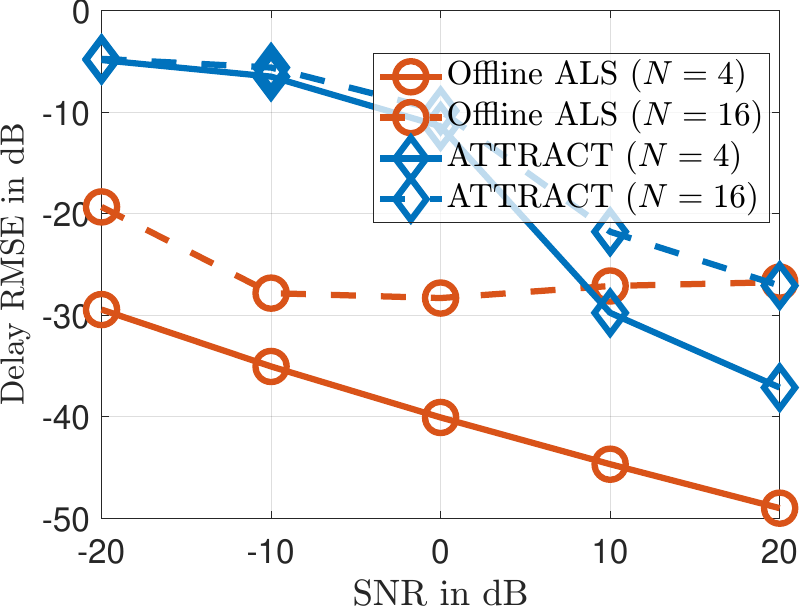}
                \caption*{(c) Delay.}
            \end{minipage}
            \hfill
            \begin{minipage}{0.475\columnwidth}   
                \centering
                \includegraphics[width=\textwidth]{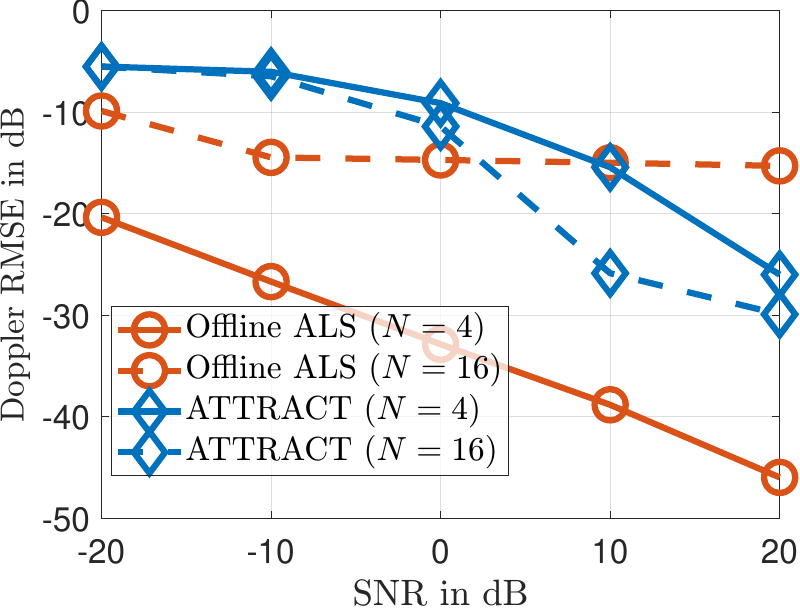}
                \caption*{(d) Doppler.}  
            \end{minipage}
            \caption{\ac{RMSE} comparison between \ac{ATTRACT} and the offline \ac{ALS} baseline for channel, angle, delay, and Doppler estimation.}
            \label{fig:rmse_case_1}
        \end{figure}

        Figure~\ref{fig:rmse_case_1} compares the \ac{RMSE} of \ac{ATTRACT} with that of the offline \ac{ALS} baseline. The offline estimator benefits from access to the full sensing window, whereas \ac{ATTRACT} updates its estimates sequentially. Despite this difference, the accuracy gap narrows as the \ac{SNR} approaches $20$~dB. In the evaluated configurations, increasing the number of reflecting elements degrades the offline baseline's accuracy but improves that of \ac{ATTRACT}. The representation in (\ref{eq:PARAFAC}) contains $N^{2}$ components, making unstructured factor estimation more demanding as $N$ increases. By reusing preceding estimates and imposing the known delay and Doppler structure, \ac{ATTRACT} provides a more favorable accuracy--complexity tradeoff. 
        
\begin{figure}[!t]
            \centering
            \setlength{\abovecaptionskip}{3pt}
            \setlength{\belowcaptionskip}{-1pt}
            \begin{minipage}{0.48\columnwidth}
                \centering
                \includegraphics[width=\textwidth]{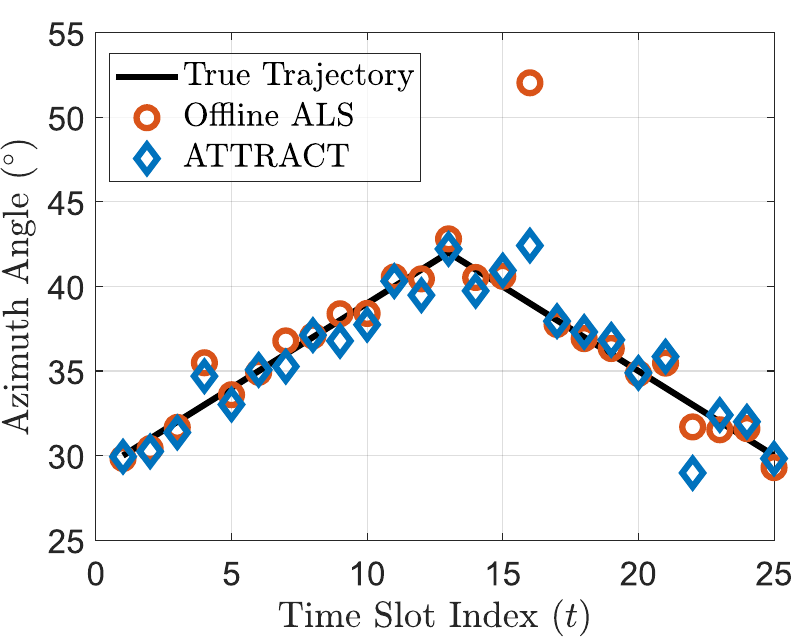}
                \caption*{(a) Azimuth tracking.}
            \end{minipage}
            \hfill
            \begin{minipage}{0.48\columnwidth}  
                \centering
                \includegraphics[width=\textwidth]{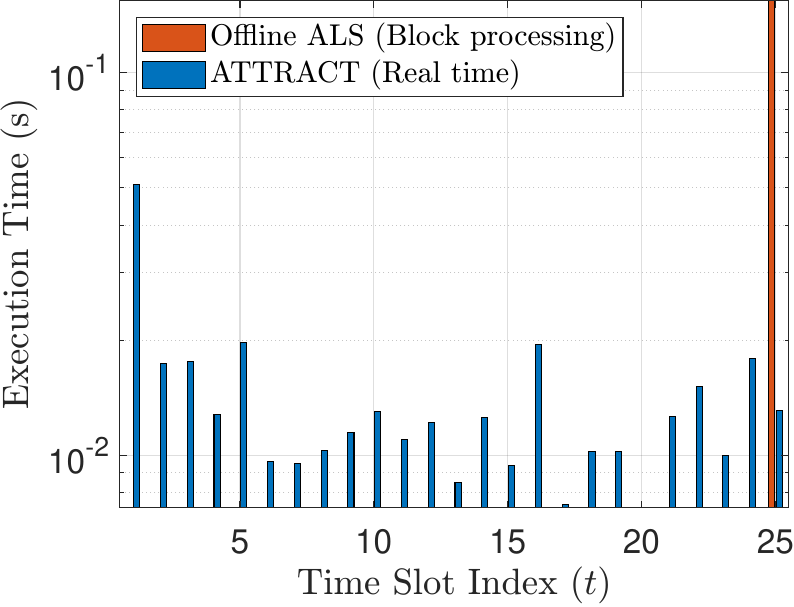}
                \caption*{(b) Execution time}
            \end{minipage}
            \caption{Angular tracking and execution-time comparison between \ac{ATTRACT} and the offline \ac{ALS} baseline ($15$~dB \ac{SNR}, $N = 4$).}
            \label{fig:latency_case_1}
        \end{figure}

        Figure~\ref{fig:latency_case_1} examines angular tracking and execution time at an \ac{SNR} of $15$~dB with $N = 4$. In Fig.~\ref{fig:latency_case_1}(a), a step change in the azimuth trend at $t = 13$ tests the response to a directional shift. The offline \ac{ALS} estimates exhibit outliers, including at $t = 16$, whereas \ac{ATTRACT} follows the trajectory more closely in this example. Figure~\ref{fig:latency_case_1}(b) shows an execution time on the order of $10^{-2}$~s per slot for \ac{ATTRACT}. Beyond this computational benefit, the sequential formulation removes the need to wait until the end of the sensing window before producing an updated estimate. This distinction between processing time and data-collection delay is central to the proposed low-latency tracking approach.
    \section{Conclusion}
        This paper has introduced \ac{ATTRACT}, a structure-aware tensor algorithm for low-latency angular tracking in \ac{RIS}-assisted sensing. Its central contribution is integrating slice-wise adaptive \ac{ALS} with delay- and Doppler-factor refinement based on the known sensing configuration. By carrying forward factor estimates and sequentially processing each new observation slice, \ac{ATTRACT} updates the target state without assembling the full sensing window or requiring an explicit motion model. Numerical results demonstrate an attractive accuracy--latency tradeoff: \ac{ATTRACT} approaches the offline baseline's accuracy at high \ac{SNR} while achieving execution times on the order of $10^{-2}$~s per slot in the evaluated setup. Beyond reducing computation, its sequential operation avoids the data-collection delay associated with a complete sensing window. These findings support structure-aware estimation as a practical approach to responsive \ac{RIS}-assisted sensing. Future work will examine robustness to imperfect \ac{ST}--\ac{RIS} channel estimates, \textcolor{black}{identifiability and uniqueness properties of \ac{ATTRACT}}, and extend the framework to multiple targets with time-varying delay and Doppler.
    
    
    \bibliographystyle{IEEEtran}
    \bibliography{IEEEexample}

\begin{thebibliography}{10}
\providecommand{\url}[1]{#1}
\csname url@samestyle\endcsname
\providecommand{\newblock}{\relax}
\providecommand{\bibinfo}[2]{#2}
\providecommand{\BIBentrySTDinterwordspacing}{\spaceskip=0pt\relax}
\providecommand{\BIBentryALTinterwordstretchfactor}{4}
\providecommand{\BIBentryALTinterwordspacing}{\spaceskip=\fontdimen2\font plus
\BIBentryALTinterwordstretchfactor\fontdimen3\font minus
  \fontdimen4\font\relax}
\providecommand{\BIBforeignlanguage}[2]{{%
\expandafter\ifx\csname l@#1\endcsname\relax
\typeout{** WARNING: IEEEtran.bst: No hyphenation pattern has been}%
\typeout{** loaded for the language `#1'. Using the pattern for}%
\typeout{** the default language instead.}%
\else
\language=\csname l@#1\endcsname
\fi
#2}}
\providecommand{\BIBdecl}{\relax}
\BIBdecl

\bibitem{liu2022integrated}
F.~Liu, Y.~Cui, C.~Masouros, J.~Xu, T.~X. Han, Y.~C. Eldar, and S.~Buzzi,
  ``Integrated sensing and communications: Toward dual-functional wireless
  networks for {6G} and beyond,'' \emph{IEEE Journal on Selected Areas in
  Communications}, vol.~40, no.~6, pp. 1728--1767, 2022.

\bibitem{zhang2025intelligent}
J.~Zhang, W.~Lu, C.~Xing, N.~Zhao, N.~Al-Dhahir, G.~K. Karagiannidis, and
  X.~Yang, ``Intelligent integrated sensing and communication: A survey,''
  \emph{Science China Information Sciences}, vol.~68, no.~3, p. 131301, 2025.

\bibitem{benicio2026multi}
K.~Ben{\'\i}cio, A.~L. de~Almeida, F.-E. Asim, B.~Sokal, G.~Fodor, B.~Makki,
  and A.~L. Swindlehurst, ``Multi-target estimation via tensor decomposition
  for beyond diagonal {RIS}-aided bistatic sensing,'' \emph{arXiv preprint
  arXiv:2604.18852}, 2026.

\bibitem{chepuri2023integrated}
S.~P. Chepuri, N.~Shlezinger, F.~Liu, G.~C. Alexandropoulos, S.~Buzzi, and
  Y.~C. Eldar, ``Integrated sensing and communications with reconfigurable
  intelligent surfaces: From signal modeling to processing,'' \emph{IEEE Signal
  Processing Magazine}, vol.~40, no.~6, pp. 41--62, 2023.

\bibitem{de2021channel}
G.~T. de~Ara{\'u}jo, A.~L. de~Almeida, and R.~Boyer, ``Channel estimation for
  intelligent reflecting surface assisted {MIMO} systems: A tensor modeling
  approach,'' \emph{IEEE J. Sel. Top. Signal Process}, vol.~15, no.~3, pp.
  789--802, 2021.

\bibitem{trice2021}
K.~Ardah, S.~Gherekhloo, A.~L.~F. de~Almeida, and M.~Haardt, ``Trice: A channel
  estimation framework for {RIS}-aided millimeter-wave {MIMO} systems,''
  \emph{IEEE Signal Processing Letters}, vol.~28, pp. 513--517, 2021.

\bibitem{gong2020toward}
S.~Gong, X.~Lu, D.~T. Hoang, D.~Niyato, L.~Shu, D.~I. Kim, and Y.-C. Liang,
  ``Toward smart wireless communications via intelligent reflecting surfaces: A
  contemporary survey,'' \emph{IEEE Commun. Surv. Tutor.}, vol.~22, no.~4, pp.
  2283--2314, 2020.

\bibitem{Swindle2022}
A.~L. Swindlehurst, G.~Zhou, R.~Liu, C.~Pan, and M.~Li, ``Channel estimation
  with reconfigurable intelligent surfaces—{A} general framework,''
  \emph{Proceedings of the IEEE}, vol. 110, no.~9, pp. 1312--1338, 2022.

\bibitem{benicio2023tensor}
K.~B. Ben{\'\i}cio, A.~L. de~Almeida, B.~Sokal, F.~E-Asim, B.~Makki, and
  G.~Fodor, ``Tensor-based channel estimation and data-aided tracking in
  {IRS}-assisted {MIMO} systems,'' \emph{IEEE Wireless Communications Letters},
  vol.~13, no.~2, pp. 333--337, 2023.

\bibitem{benicio2024low}
K.~B. Ben{\'\i}cio, B.~Sokal, A.~L. de~Almeida, B.~Makki, G.~Fodor
  \emph{et~al.}, ``Low-complexity tensor-based monostatic sensing for
  {IRS}-assisted communication systems,'' in \emph{2024 19th International
  Symposium on Wireless Communication Systems (ISWCS)}.\hskip 1em plus 0.5em
  minus 0.4em\relax IEEE, 2024, pp. 1--6.

\bibitem{benicio2024ris}
K.~Ben{\'\i}cio, B.~Sokal, A.~L. de~Almeida, B.~Makki, G.~Fodor, A.~L.
  Swindlehurst \emph{et~al.}, ``{RIS}-assisted sensing: A nested tensor
  decomposition-based approach,'' in \emph{Proc. 58th Asilomar Conference on
  Signals, Systems, and Computers}, 2024, pp. 1581--1585.

\bibitem{comon2009tensor}
P.~Comon, X.~Luciani, and A.~L. De~Almeida, ``Tensor decompositions,
  alternating least squares and other tales,'' \emph{Journal of Chemometrics: A
  Journal of the Chemometrics Society}, vol.~23, no. 7-8, pp. 393--405, 2009.

\bibitem{abed2022contemporary}
L.~T. Thanh, K.~Abed-Meraim, N.~L. Trung, and A.~Hafiane, ``A contemporary and
  comprehensive survey on streaming tensor decomposition,'' \emph{IEEE
  Transactions on Knowledge and Data Engineering}, vol.~35, no.~11, pp.
  10\,897--10\,921, 2022.

\bibitem{nion2009adaptive}
D.~Nion and N.~D. Sidiropoulos, ``Adaptive algorithms to track the parafac
  decomposition of a third-order tensor,'' \emph{IEEE Transactions on Signal
  Processing}, vol.~57, no.~6, pp. 2299--2310, 2009.

\bibitem{Asim_2025}
Fazal-E-Asim, A.~L.~F. de~Almeida, B.~Sokal, B.~Makki, and G.~Fodor,
  ``Two-dimensional channel parameter estimation for {IRS}-assisted networks,''
  \emph{IEEE Transactions on Communications}, vol.~73, no.~8, pp. 6337--6350,
  2025.

\bibitem{ercan2025ris}
M.~K. Ercan, A.~Pourafzal, M.~F. Keskin, S.~Gezici, and H.~Wymeersch,
  ``{RIS}-aided {NLoS} monostatic multi-target sensing under angle-{D}oppler
  coupling,'' \emph{IEEE Trans. Veh. Technol.}, vol.~74, no.~12, pp. 19\,141 --
  19\,158, 2025.

\bibitem{zhang2016tracking}
C.~Zhang, D.~Guo, and P.~Fan, ``Tracking angles of departure and arrival in a
  mobile millimeter wave channel,'' in \emph{IEEE International Conference on
  Communications (ICC)}, 2016.

\end{thebibliography}
    
\end{document}